\documentclass[reprint,prx,superscriptaddress,longbibliography,nofootinbib]{revtex4-1}
\usepackage{amsmath}
\usepackage{amssymb}
\usepackage{graphicx}
\usepackage{bm}
\usepackage{times}
\usepackage{blkarray}
\usepackage{multirow}
\usepackage{color}
\usepackage{stmaryrd}

\usepackage{enumerate}
\usepackage{graphicx}
\usepackage{amsmath}
\usepackage{amssymb}
\usepackage{amsthm}
\usepackage{amscd}
\usepackage{amsfonts}
\usepackage{makeidx}
\usepackage{bm}
\usepackage{tensor}
\usepackage{eucal}
\usepackage{epsfig}

\def\la{\langle}
\def\ra{\rangle}

\newcommand{\mean}[1]{\la#1\ra}                     
\newcommand{\ket}[1]{\vert#1\ra}                    

\newcommand{\nn}{\nonumber}

\DeclareMathOperator{\env}{Env}
\DeclareMathOperator{\sinc}{Sinc}

\begin{document}

\title{Compressively characterizing high-dimensional entangled states
  with complementary, random filtering}


\author{Gregory A. Howland}
\email{gregory.howland.3@us.af.mil}
\affiliation{Department of Physics and Astronomy, University of Rochester \\
500 Wilson Blvd, Rochester, NY 14627, USA}
\affiliation{Air Force Research Laboratory\\ 525 Brooks Rd, Rome, NY 13441}
\author{Samuel H. Knarr}
\email{sknarr@ur.rochester.edu}
\affiliation{Department of Physics and Astronomy, University of Rochester \\
500 Wilson Blvd, Rochester, NY 14627, USA}
\author{James Schneeloch}
\affiliation{Department of Physics and Astronomy, University of Rochester \\
500 Wilson Blvd, Rochester, NY 14627, USA}
\affiliation{Air Force Research Laboratory\\ 525 Brooks Rd, Rome, NY 13441}
\author{Daniel J. Lum}
\author{John C. Howell}
\affiliation{Department of Physics and Astronomy, University of Rochester \\
500 Wilson Blvd, Rochester, NY 14627, USA}

\begin{abstract}
  The resources needed to conventionally characterize a quantum system
  are overwhelmingly large for high-dimensional systems.  This
  {\color{black}obstacle} may be overcome by abandoning traditional
  cornerstones of quantum measurement, such as general quantum states,
  strong projective measurement, and assumption-free
  characterization. Following this reasoning, we demonstrate an
  efficient technique for characterizing high-dimensional, spatial
  entanglement {\color{black}with one set of measurements}. We recover
  sharp distributions with local, random filtering of the same
  ensemble in momentum followed by position---something the
  uncertainty principle forbids for projective
  measurements. Exploiting the expectation that entangled signals are
  highly correlated, we use fewer than $5,000$ measurements to
  characterize a $65,536$-dimensional state. Finally, we use entropic
  inequalities to witness entanglement without a density matrix. Our
  method represents the sea change unfolding in quantum measurement
  where methods influenced by the information theory and
  signal-processing communities replace unscalable, brute-force
  techniques---a progression previously followed by classical sensing.
\end{abstract}

\maketitle
\newcommand\blfootnote[1]{%
  \begingroup
  \renewcommand\thefootnote{}\footnote{#1}%
  \addtocounter{footnote}{-1}%
  \endgroup
}

\blfootnote{G.A.H. and S.H.K contributed equally to this work.}

Practicing experimentalists most commonly perform quantum measurement
in the context of state and parameter estimation
\cite{paris:2004}. While great historical emphasis has been placed on
using measurement to probe the validity of quantum mechanics
itself---where measurements must not only agree with quantum
predictions but also rule out any competing explanations
\cite{zeilinger:1998}---state estimation accepts quantum theory
\emph{a priori}. Here, measurements on identically prepared copies of
a system are used to generate a model from which \emph{testable}
predictions can be made about future measurement statistics
\cite{rozema:2014}.  This point of view lifts the burden of
validation, leading to simpler experiments {\color{black} and
  technologies.}

Even so, quantum state estimation remains a persistent obstacle for
scaling quantum technologies. The familiar approach of quantum
tomography (QT) scales at least quadratically poorly with added
dimensions and exponentially poorly with added particles. QT in an
$N$-dimensional Hilbert space requires of order $N^2$ measurements
\cite{thew:2002}---when $N$ is a prime power, $N$ projections are
taken in each of $N+1$ mutually unbiased bases
\cite{klappenecker:2004}. For example, tomography of a single spin
qubit ($N=2$) requires dividing the ensemble three ways, where
expectation values of the $\hat{X}$, $\hat{Y}$ and $\hat{Z}$ spin
components are separately measured. For most non-trivial quantum
systems, traditional, brute-force QT is unmanageable in the lab. In
particular, continuous-variable degrees-of-freedom, such as
transverse-position and transverse-momentum or energy and time, where
$N \rightarrow \infty$, cannot be realistically characterized via QT
\cite{lvonsky:2009}.

Efforts to overcome the limitations of QT fall into three major
categories. First, often only a subset of a system's behavior is of
interest; e.g., if one only needs to predict a qubit's spin along one
axis, information about the other two is irrelevant. The general
tomographic density matrix can here be discarded in favor of simpler
models \cite{aaronson:2007}. A practical example is quantum key
distribution (QKD), where only two (instead of order $N$) bases, such
as energy and time, need be characterized \cite {ali-khan:2007}. Many
entanglement witnesses only require a small subset of possible
measurements to confirm entanglement
\cite{horodecki:2009,cavalcanti:2009}.

\begin{figure*}[t]
  \begin{centering}
    \includegraphics[scale=0.4]{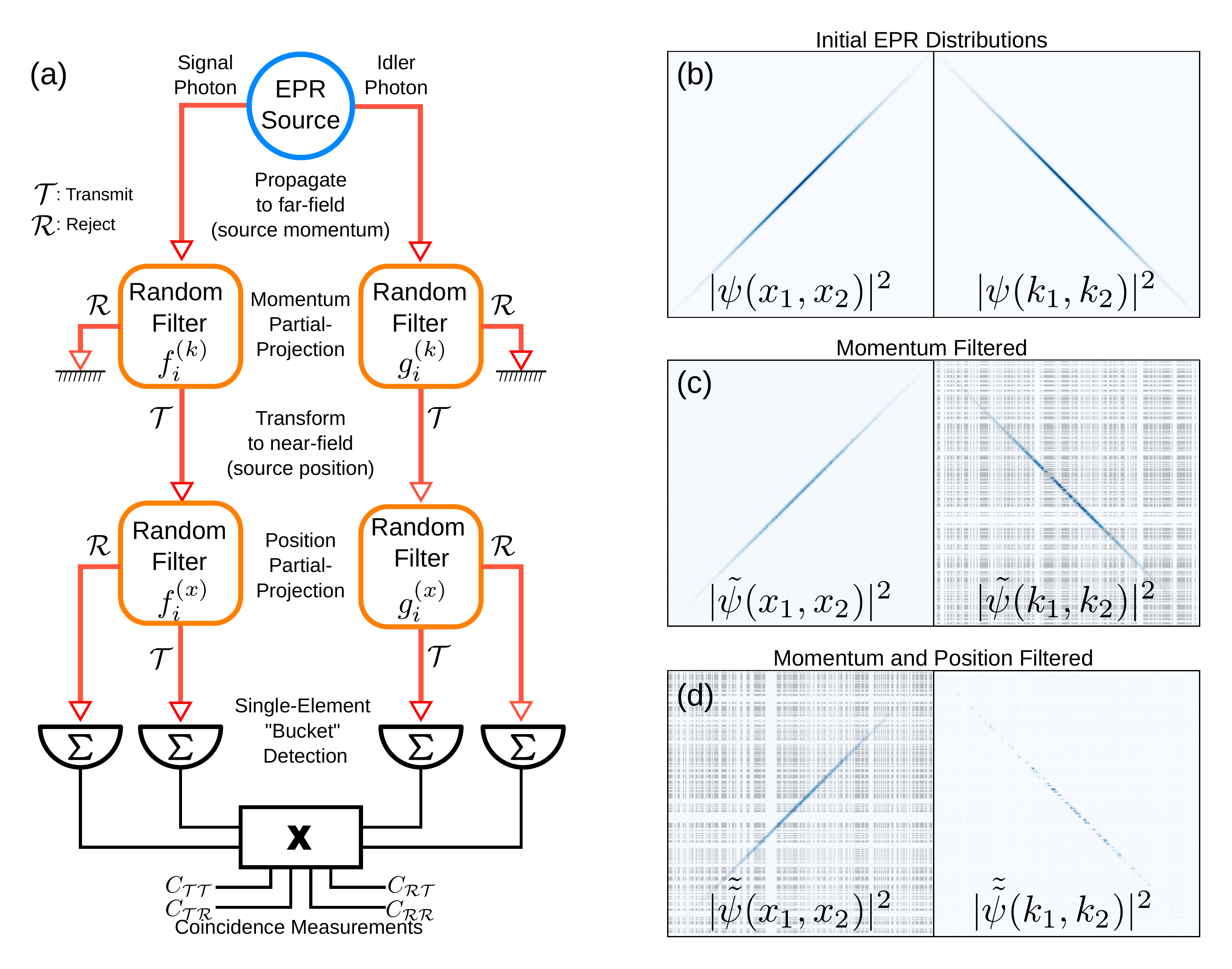}
    \caption{{\bf Sequential, partial-projections in position and
        momentum:} The block diagram (a) describes a sequence of
      partially-projective measurements on an EPR entangled
      source. (b---d) show {\color{black} simulated} joint-position and
      joint-momentum distributions at each point in the
      experiment. Signal and idler photons from an EPR source (b) are
      separated and allowed to propagate to the far-field
      (momentum). Here they are subjected to random binary filtering
      by a pixelated mask ({\color{black}faded gray overlay}). Each
      pixel in the mask either fully transmits ($\mathcal{T}$) or
      fully rejects ($\mathcal{R}$). The momentum-filtered fields (c)
      propagate through an optical system to an image plane of the
      source, where they are again filtered with random, binary
      filters (d). Single-element, photon-counting detectors are
      placed in the $\mathcal{T}$ and $\mathcal{R}$ ports of each
      filter and are connected to a coincidence circuit. The total
      number of coincident detection events between signal and idler
      channels gives a random projection of the momentum
      distribution. The relative distribution of coincident detections
      between the $\mathcal{T}$ and $\mathcal{R}$ modes (4
      possibilities) for the signal and idler photons gives a random
      projection of the position distribution up to a small noise
      floor injected by the momentum filtering.}
    \label{fig:concept}
  \end{centering}
\end{figure*}

Second, one can leverage prior knowledge about a system.  In standard
tomography, maximum likelihood estimation is used to find a valid
density matrix consistent with measurement data \cite{hradil:1997,
  james:2001}---a simple assumption that quantum mechanics holds. Or,
given a model of the physical system, one can begin with a prior
distribution which is updated or parameterized in response to
measurements, as in Bayesian inference \cite{caves:2002,huszar:2012}.

One powerful presupposition is that a signal is structured, or
\emph{compressible}. For classical signals, this surprisingly broad
assumption spawned the field of compressed sensing (CS) to tremendous
multidisciplinary impact \cite{donoho:2006,carron:2009} with a strong
presence in imaging \cite{duarte:2008, romberg:2008, sun:2013,
  kirmani:2014}. In compressed sensing, signals are compressed during
measurement so they can be sampled below the Nyquist limit
\cite{candes:2006}. Several recent efforts apply CS to quantum
measurement to dramatic effect \cite{gross:2010,shabani:2011,
  liu:2012,tonolini:2014,weinfurter:2014}---in some cases reducing
measurement times from years to hours
\cite{howland:2013}. {\color{black} For tomography, all protocols
  exploiting positivity are a form of compressed
  sensing \cite{kalev:2015}.}

Finally, one can choose measurements well suited to the model and
prior knowledge. There is a compelling movement beyond traditional,
projective measurements that localize quantum particles. Notably,
there is weak measurement, where a system and measurement device are
very weakly coupled, leaving the system nearly undisturbed
\cite{dressel:2014}. With weak measurement, researchers have directly
measured the quantum wavefunction \cite{lundeen:2011}, observed
average trajectories of particles in the double-slit experiment
\cite{kocsis:2011}, and performed tests of local realism
\cite{white:2015}. More recently, we investigated partially-projecting
measurements which lie somewhere between weak and projective
measurement. Using random, binary, filtering in position followed by
strong projections in momentum, we measured the sharp image and
diffraction pattern of a transverse optical field without dividing the
initial ensemble, a feat impossible for strong, projective
measurements \cite{howland:2014}. With non-projective measurement, the
conventional wisdom that incompatible variables must be separately
investigated is discarded.

Guided by these principles, we demonstrate a novel approach for
efficiently witnessing large-dimensional entanglement with a single
set of measurements. We apply this technique to
Einstein-Podolsky-Rosen (EPR) correlations in the spatial
degrees-of-freedom of the biphoton state produced in spontaneous
parametric down-conversion (SPDC), a system closely resembling the EPR
\emph{gedankenexperiment} \cite{epr, howell:2004}. Inspired by the
random measurements used in CS, we show that random, local,
partial-projections in momentum followed by random, local, partial
projections in position can be used to efficiently and accurately
image EPR correlations in both domains. The ensemble is not
split---position and momentum measurements are performed on the same
photons.  Remarkably, the measurement disturbance introduced by the
momentum filtering manifests as a small amount of additive noise in
the position distribution, which remains un-broadened. This allows the
position and momentum measurements to be decoupled, and the joint
probability distributions to be recovered in a $65,536$-dimensional
discretization of the infinite-dimensional Hilbert space.
{\color{black} Our measurements do not violate the uncertainty
  principle; rather, they highlight the complex and subtle behavior of
  measurement disturbance given non-projective measurements.}

Exploiting our expectation that the distributions are highly
correlated, we use compressive sensing optimization techniques to
dramatically undersample---we need fewer than $5,000$ measurements to
obtain high-quality distributions. By comparing the conditional
Shannon entropy in the position and momentum joint distributions, we
witness high-dimensional entanglement and determine a quantum secret
key rate for the joint system without needing a density matrix.

\section{Theory}
\subsection{Random, partially-projective measurements of an EPR state}

Consider a two-photon quantum state $\ket{\psi}$ encoded in the
transverse-spatial degrees of freedom of the biphoton produced by
SPDC. SPDC is a nonlinear-optical process where a high-energy pump
photon is converted into two lower-energy daughter photons, labeled
signal and idler. Conservation of momentum dictates that the signal
and idler momenta be anti-correlated for a plane wave
pump. Conservation of ``birthplace'', the notion that both photons
originate from the same location in the crystal, dictates positive
correlations in the daughters' transverse-positions.

Strong correlations in incompatible observables are a signature of
entanglement---in fact, the original EPR paradox was described using
position and momentum \cite{epr}. EPR considered the ideal state
\begin{align}
  \ket{\psi} &=\int dx_1dx_2 \delta (x_1-x_2)\ket{x_1,x_2}\\
  &=\int dk_1dk_2 \delta (k_1+k_2)\ket{k_1,k_2};\nonumber
\end{align}
perfectly correlated in position and perfectly anti-correlated in
momentum. Although the ideal EPR state is non-normalizable and
consequently impossible to realize in the lab, the biphoton state
generated via SPDC is very similar \cite{walborn:2010,
  schneeloch:2015}.

EPR correlations are observed by measuring the joint probability
distribution in position, $|\psi(x_1,x_2)|^2$, and in momentum
$|\psi(k_1,k_2)|^2$. Because these domains of interest are known in
advance, only these two distributions are needed---not a full density
matrix. Spatial correlations are usually measured by jointly raster
scanning single-element, photon-counting detectors through either the
near-field (position) or far-field (momentum) \cite{howell:2005}.
This approach scales extremely poorly with increased single-particle
dimensionality $n$--- measurement time scales between $n^3$ and
$n^4$. For a typical source, this could take upwards of one year for a
modest $n=32\times 32$ pixel resolution \cite{howland:2013}.

To avoid dividing the ensemble, and to require many fewer
measurements, we instead apply local, partially-projective
measurements in momentum followed by local, partially-projective
measurements in position, to the same photons. Our approach is
illustrated in Fig. \ref{fig:concept}. The signal and idler photons
from an EPR-like state $\psi(x_1,x_2)$ are separately allowed to
propagate to the far-field. Here, each photon is locally filtered by a
random, binary mask $f_i^{(k)}(k_1)$ (signal) or $g_i^{(k)}(k_2)$
(idler), where subscript $i$ refers to a particular pair of
filters. Each local filter is an $n$-pixel, binary intensity mask,
where individual pixels fully transmit ($\mathcal{T}$) or fully reject
($\mathcal{R}$) with equal probability. The momentum filtering enacts
a significant partial-projection of $\ket{\psi}$---on average, half of
the local intensity and three-quarters of the joint-intensity is
rejected---{\color{black}so this is not a weak measurement}.

{\color{black} All measurements are subject to uncertainty relations,
  which imply unavoidable measurement disturbance. Conventional
  projective measurements, often associated with ``wavefunction
  collapse'', localize a quantum state in one domain (e.g momentum) at
  the cost of broadening it in a conjugate domain (e.g. position).}
Critically, however, {\color{black} random filtering} \emph{does not
  localize} the quantum state; it maps a small amount of momentum
information onto the total intensity passing the filter. The
measurement disturbance of non-projective measurements is best
understood via the entropic uncertainty principle
\begin{equation}
  \label{eq:ent}
  h(x) + h(k) \ge \log(\pi e),
\end{equation}
where $h(*)$ is the Shannon entropy. The entropic uncertainty
principle implies an information exclusion relation; \emph{the more
  information a measurement gives about the momentum distribution, the
  less information a subsequent measurement can give about the
  position distribution} \cite{hall:1995}. There are no restrictions,
however, on how information loss manifests. {\color{black} In
  particular, a measurement in one domain need not broaden, or blur,
  the statistics in a complementary domain.}

The joint amplitude passing the momentum-filtering is
$\tilde{\psi}(k_1,k_2) = \psi(k_1,k_2)f_i^{(k)}(k_1)g_i^{(k)}(k_2)$.
To see the effect of the momentum filtering on the position
distribution, we take a Fourier transform to find
$\tilde{\psi}(x_1,x_2) =
\mathcal{F}\left\{\tilde{\psi}(k_1,k_2)\right\} $,
which is given by the convolution of the state and filter functions in
the position domain:
$\tilde{\psi}(x_1,x_2) = \psi(x_1,x_2)\star
(f_i^{(k)}(x_1)g_i^{(k)}(x_2))$.
At high resolution, the Fourier transform of an $n$-pixel, random
binary pattern is approximately proportional to
$\delta(x) + \sqrt{2/n}\;\phi(x)$, where values for $\phi(x)$ are
taken from a unit variance, complex, Gaussian noise distribution---a
sharp central peak riding a small noise floor \cite{howland:2014} (see
supplemental material).

Because convolution with a delta function returns the original
function, the perturbed state's position distribution is the true
distribution with some weak additive noise terms;
\begin{align}
|\tilde{\psi}&(x_1,x_2)|^{2}=\mathcal{N}\bigg|\psi(x_1,x_2)\star
\\&\left[\left(\delta(x_1) + \sqrt{2/N}\phi_i(x_1)\right)\left(\delta(x_2)+\sqrt{2/N}\phi_i(x_2)
\right)\right]\bigg|^2\nonumber
\end{align}
Expanding this product in powers of $1/\sqrt{N}$, where $N=n^2$,
yields
\begin{align}
  \label{eq:perteffect}
  |\tilde{\psi}&(x_1,x_2)|^{2} = \mathcal{N}\biggl\{|\psi(x_1,x_2)|^2\\
                              &+ \sqrt{2/N}\textbf{Re}\bigg[\psi^*(x_1,x_2)\big(\psi(x_1,x_2)\star\nonumber\\
&\phantom{qqqqq}(\delta(x_1)\phi_2(x_2)+\delta(x_2)\phi_1(x_1))\big)\bigg] \nonumber\\
                              &+ \mathcal{O}(1/N) + \dots + \mathcal{O}(1/N^2)\biggr\}  \nonumber
\end{align}
where $\mathcal{N}$ is a normalizing constant. Remarkably, disturbance
from filtering adds only a small noise floor at most a factor
$\sqrt{2/N}$ weaker without otherwise broadening the position
distribution. This can be seen in Fig. \ref{fig:concept}(c), where the
position distribution maintains tight correlations despite the effect
of momentum filtering. A rigorous derivation of equation
(\ref{eq:perteffect}), including the effect of finite width pixels, is
given in the supplemental material.

Next, we again perform random filtering---this time in position---as
seen in Fig. \ref{fig:concept}(d). The transmitted and rejected ports
are directed to single-element ``bucket'' detectors that are not
spatially resolving. Photon detection events are time-correlated with
a coincidence circuit.

Each coincidence measurement contains information about both position
and momentum; these must be decoupled to fit a measurement model
\begin{align}   \label{eq:meas}
   \bm{Y}^{(k)} &= \bm{AK} + \bm{\Phi}^{(k)}\\
   \bm{Y}^{(x)} &= \bm{BX} + \bm{\Phi}^{(x)} + \bm{\Gamma}^{(x)}. \nonumber
\end{align}
Here, $\bm{K}$ and $\bm{X}$ are $N$-dimensional signal vectors
representing $|\psi(k_1,k_2)|^2$ and $|\psi(x_1,x_2)|^2$, and $\bm{A}$
and $\bm{B}$ are $M \times N$ sensing matrices. $\bm{Y}^{(k)}$ and
$\bm{Y}^{(x)}$ are measurement vectors whose elements are the
inner-product of $\bm{X}$ or $\bm{K}$ onto the $i^{\text{th}}$ row (or
sensing vector) of $\bm{A}$ or $\bm{B}$. Noise vectors $\bm{\Phi}$
represent additive measurement noise. Noise vector $\bm{\Gamma}^{(x)}$
represents the noise injected by momentum filtering.

Momentum information is encoded in the \emph{total} coincidences
between all detection modes. Each row of $\bm{A}$ is the Kronecker
product of two, random single-particle sensing vectors
$\bm{a}^{k_1}_i\otimes \bm{a}^{k_2}_i$ such that
$\bm{A}_i = \bm{a}_i^{k_1}\otimes \bm{a}_i^{k_2}$, where for example
$\bm{a}_i^{k_1}$ encodes $f_i^{(k)}(k_1)$.

\begin{figure*}[t]
  \begin{centering}
    \includegraphics[scale=0.2]{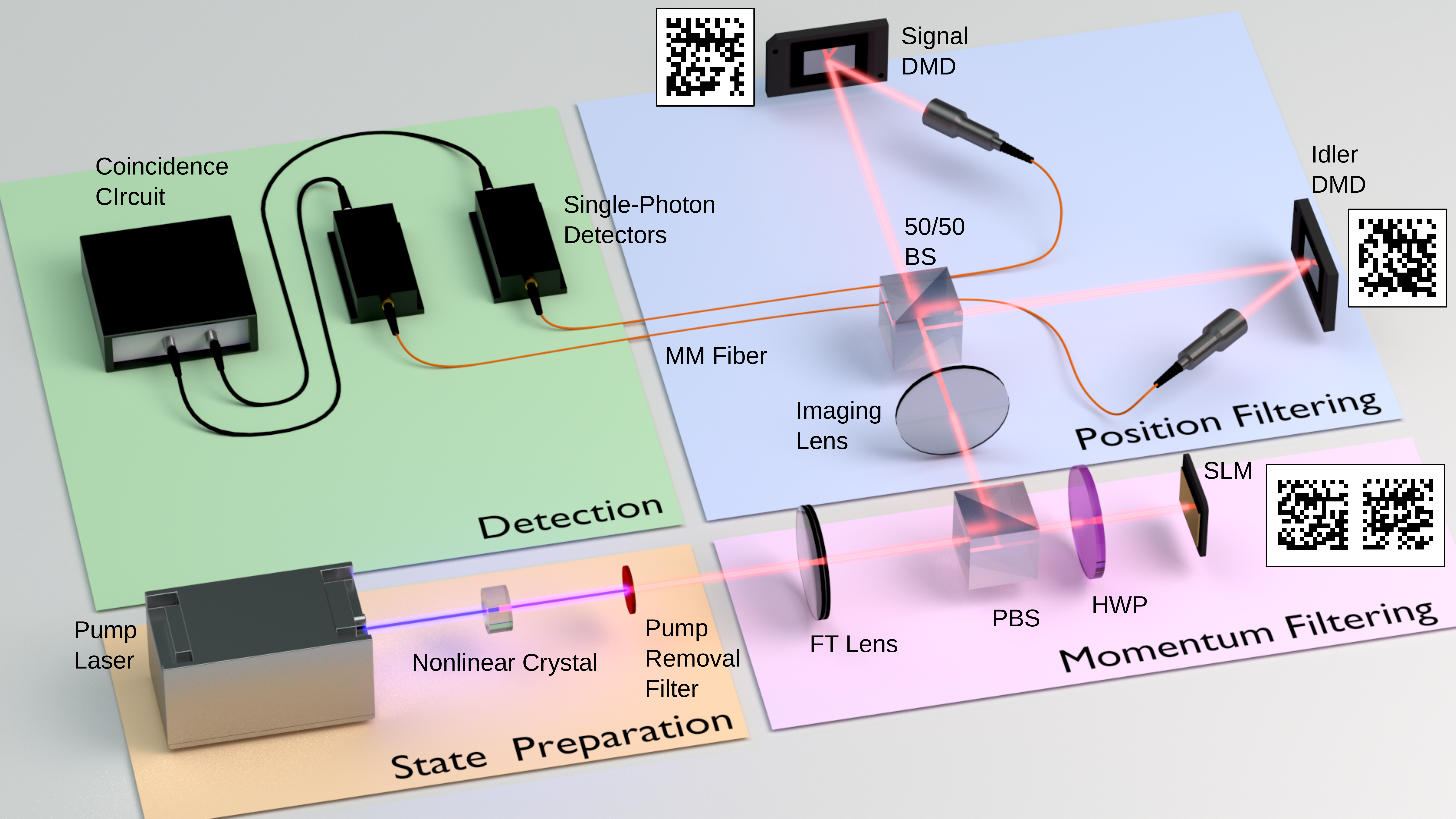}
    \caption{{\bf Experimental Setup:} A two-photon, EPR-like state is
      generated by pumping a nonlinear crystal for Type-1
      SPDC. Random, binary patterns placed on an SLM in a Fourier
      plane of the crystal and on DMDs in an image plane of the
      crystal implement a sequence of random, partially-projecting
      measurements. {\color{black} Example patterns are shown next to
        the SLM and DMDs; note the separate patterns for signal and
        idler photons on the SLM.} Coincident detection events between
      single-photon detectors for signal and idler photons give
      information about \emph{both} the joint-position and
      joint-momentum distributions of the two-photon state.}
    \label{fig:setup}
  \end{centering}
\end{figure*}

Position information is encoded in the \emph{relative} distribution of
coincidences between signal and idler $\mathcal{T}$ and $\mathcal{R}$
modes. By adding coincidences between like-modes ($\mathcal{TT}$ and
$\mathcal{RR}$) and subtracting coincidences between differing modes
($\mathcal{TR}$ and $\mathcal{RT}$), the effect of momentum filtering
is removed up to injected noise. Like momentum, the position sensing
vector is a Kronecker product of two local sensing vectors;
$\bm{B}_i = \bm{b}^{(x_1)}_i\otimes\bm{b}^{(x_2)}_i$. However, because
of the relative measurement, the local sensing matrices take values
``1'' for transmitting pixels and ``-1'' for rejecting pixels.

In our experiment, we use a slightly more sophisticated, but
conceptually similar, approach (see supplemental material at \emph{to
  be inserted}) that retains the transmission and rejection modes from
both momentum and position. In this case, there are 16 possible
correlation measurements that are combined to give either position or
momentum information, and both $\bm{A}$ and $\bm{B}$ take values ``1''
and ``-1''.

\subsection{Recovering the position and momentum distributions}

To obtain the joint-position and joint-momentum distributions from our
measurements, we turn to compressive sensing (CS). Here, we exploit
our expectation that both distributions are highly
correlated. Therefore, the distributions are sparse {\color{black} in
  their natural (position-pixel or momentum-pixel)
  representations}---relatively few elements in each distribution have
significant values. This allows us to dramatically under-sample so
that $M<<N$. In this case, there are many possible $\bm{X}$ and
$\bm{K}$ consistent with the measurements. CS posits that the correct
$\bm{X}$ and $\bm{K}$ are the sparsest distributions consistent with
the measurements.

Sparse $\bm{X}$ and $\bm{K}$ are found by solving a pair of
optimization problems
\begin{align}
  \min_{\bm{K}} \phantom{x}&\frac{\mu_k}{2}||\bm{Y}^{(k)}-\bm{AK}||^2_2 + TV(\bm{K}) \label{eq:opt}\\
  \min_{\bm{X}} \phantom{x}&\frac{\mu_x}{2}||\bm{Y}^{(x)}-\bm{BX}||^2_2 + TV(\bm{X}),\nonumber
\end{align}
where $||*||_2^2$ is the $\ell_2$ (Euclidean) norm, and $\mu$ are
weighting constants. The first penalty is a least-squares term that
ensures the result is consistent with measured data. The second
penalty, $TV(*)$, is the signal's total variation (TV), which is the
$\ell_1$ norm of the discrete gradient
\begin{equation}
  TV(\bm{X}) = \sum\limits_{\text{adj.}\phantom{i} i,j}|\bm{X}_i - \bm{X}_j|,
  \label{eq:tv}
\end{equation}
where $i,j$ run over pairs of adjacent elements in the signal. The TV
regularization promotes structured, sparse signals over noisy,
uncorrelated signals. Total variation minimization has been extremely
successful for compressed sensing and denoising in the context of
imaging \cite{shu:2010, chambolle:1997, li:2009}. In many cases, a
signal can be recovered from $M$ as low as a few percent of
$N$. {\color{black} For a more complete introduction to compressive
  sensing, see excellent tutorials by Baraniuk \cite{baraniuk:2007}
  and Cand{\`e}s and Wakin \cite{candes:2008}.}

Total variation minimization is also extremely effective for denoising
signals \cite{rudin:1992}. Normally, this helpfully mitigates
environmental and photon-counting shot noise ($\bm{\Phi)$}, but in our
case also largely removes the filtering measurement disturbance
$\bm{\Gamma}$. With strong measurements, e.g. raster-scanning a
pinhole aperture, one requires deconvolution techniques to obtain a
similar effect. Not only is deconvolution far more challenging than
denoising, it can never recover high frequency content beyond the
aperture size.

CS measurements are most effective in a representation that is
incoherent, or maximally unbiased, with respect to the sparse
representations (in our case position or momentum). Fortunately,
random projections perfectly suit this criteria, leading to the
surprising conclusion that random measurement is actually
preferable. Random matrices are overwhelmingly likely to be restricted
isometries that preserve the relative distance between sparse signals,
ensuring that solving Eq. \ref{eq:opt} returns the true signal instead
of a sparse but otherwise incorrect result \cite{candes:2008:2}. Not
only do random filters extract information in complementary domains,
they are the among the best measurements for leveraging CS.

{\color{black} One might reasonably ask if our technique employs
  circular reasoning---assuming the distributions are highly
  correlated in order to then measure their correlations. This is not
  the case. The initial assumption is a compressibility assumption;
  relative to all possible distributions, our distributions are
  expected to be sparse in the natural pixel basis. We do not know
  exactly how sparse the distributions will be, or which elements will
  be significant. However, the vast majority of possible distributions
  are just unstructured noise---these are the outcomes we are
  initially rejecting.

  The assumption is similar to assuming a digital photograph can be
  effectively compressed by the JPEG standard \cite{wallace:1991}. A
  natural photographic scene contains more low-spatial-frequency
  content than high-spatial-frequency content and contains objects
  with well defined edges and recognizable shapes---regardless of the
  specific scene.}

\section{Experiment}

\begin{figure*}
  \includegraphics[scale=0.6]{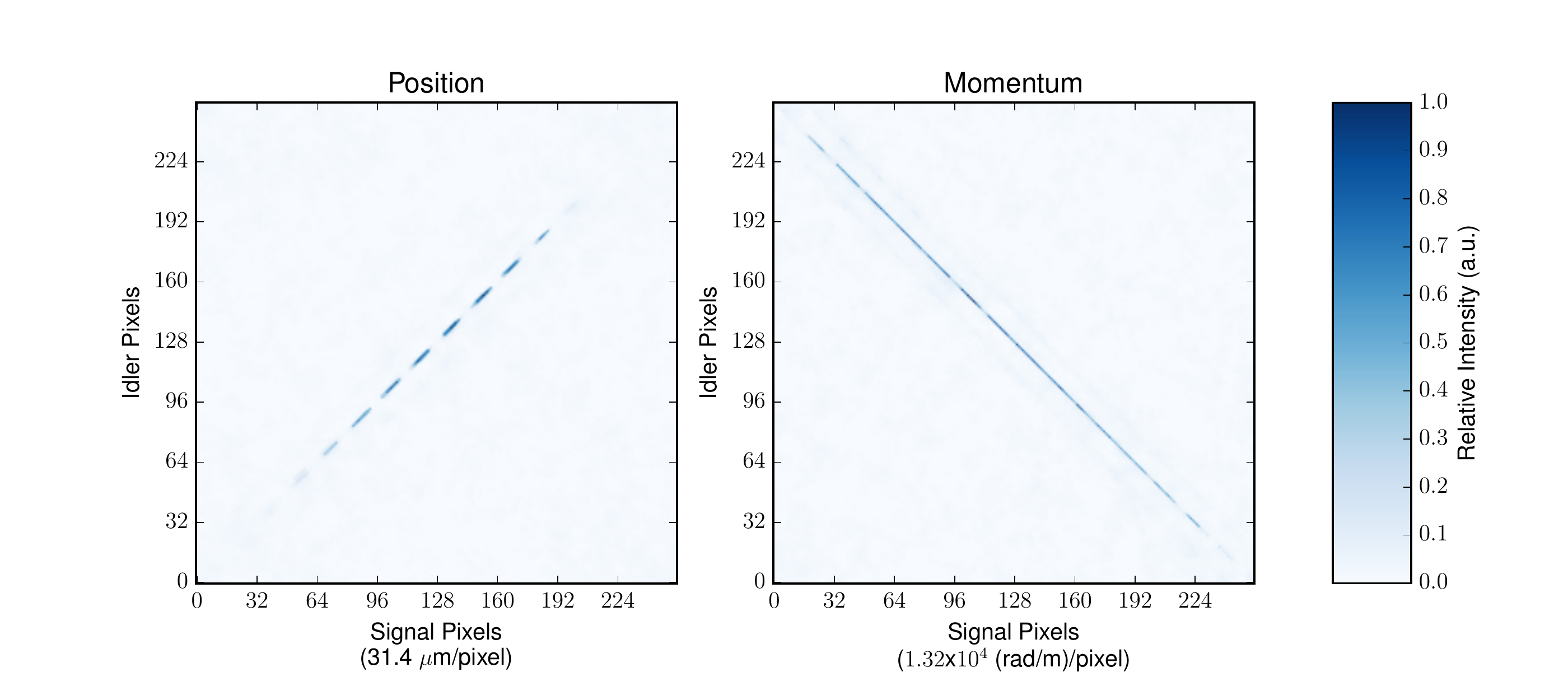}
  \caption{{\bf Representative recovered joint-position and
      joint-momentum signals.} Recovered joint-position and
    joint-momentum {\color{black}signals} for a $16\times 16$ pixel
    ($N=256\times 256$) discretization. Only $M=4,439$ measurements
    were needed, about $0.07N$. Gaps along the position diagonal occur
    due to reshaping to one dimension---these regions were outside the
    marginal width. {\color{black} Position and momentum units refer to
      the transverse plane at the nonlinear crystal ($z=0$).}}
  \label{fig:recon}
\end{figure*}

Our experimental setup is shown in Fig.  \ref{fig:setup}. An EPR-like
state at $810$ nm is generated by pumping a $1$ mm thick BiBO crystal
oriented for Type-I, collinear SPDC with a $405$ nm pump laser. The
generated fields propagate to a spatial light modulator (SLM) in the
focal plane of a $125$ mm lens.  Because the phase-only SLM only
retards one polarization, it can perform per-pixel polarization
rotation. These polarization rotations are converted to intensity
modulations with a half-wave plate and a polarizing
beamsplitter. Random masks which cause zero or $\pi$ polarization
rotations perform the momentum filtering. We exploit the negative
correlations in the momentum state to assign signal and idler
particles to the left- and right-halves of the SLM respectively.

The signal and idler fields are routed to separate digital micromirror
devices (DMDs) via a $500$ mm lens and 50/50 beamsplitter; the DMDs
are placed in a crystal image plane with $4$X magnification. A DMD is
a two-dimensional array of individually addressable mirrors, each of
which can be oriented to direct light towards or away from a
detector. These correspond to the transmit and reject ports in
Fig. \ref{fig:concept}. Random patterns placed on the DMDs implement
the position filtering. The light is coupled with $10$X microscope
objectives into multi-mode fibers which are connected to avalanche
photo-diodes operating in Geiger (photon-counting) mode. A correlator
records coincident detection events between filtered signal and idler
photons.

Single-particle sensing matrices $\bm{a}^{(k_1)}$, $\bm{a}^{(k_2)}$,
$\bm{b}^{(x_1)}$, and $\bm{b}^{(x_2)}$ are generated by taking $M$
rows from randomly permuted $n\times n$ Hadamard matrices. This allows
the repeated calculations of $\bm{AK}$ and $\bm{BX}$ performed by the
solver to use a Fast Hadamard transform, decreasing computational
requirements \cite{lum:2015}.  Because we only collect transmitted
modes from both position and momentum filters, we require $16$
separate measurements to collect all coincident combinations of
transmission and rejection for the $4$ filters (described in
supplemental material). This is not required in principle if one has
$8$ detectors. The solver we use for equation (\ref{eq:opt}) is TVAL3
\cite{li:2009:user}. {\color{black} The full measurement and
  reconstruction recipe we follow is similar to that described in
  Ref. \cite{lum:2015}.}

{\color{black} Note that our choice of a single momentum SLM and two
  position DMDs was due to available equipment. One would ideally use
  four SLMs to implement completely separate position and momentum
  filtering for both the signal and idler fields. The SLM is preferred
  for filtering because of its high ($>90\%$) diffraction efficiency
  in contrast to the lower ($\approx 20\%$) diffraction efficiency for
  the DMDs.}

\section{Results}

\begin{figure*}[t]
  \includegraphics[scale=0.5]{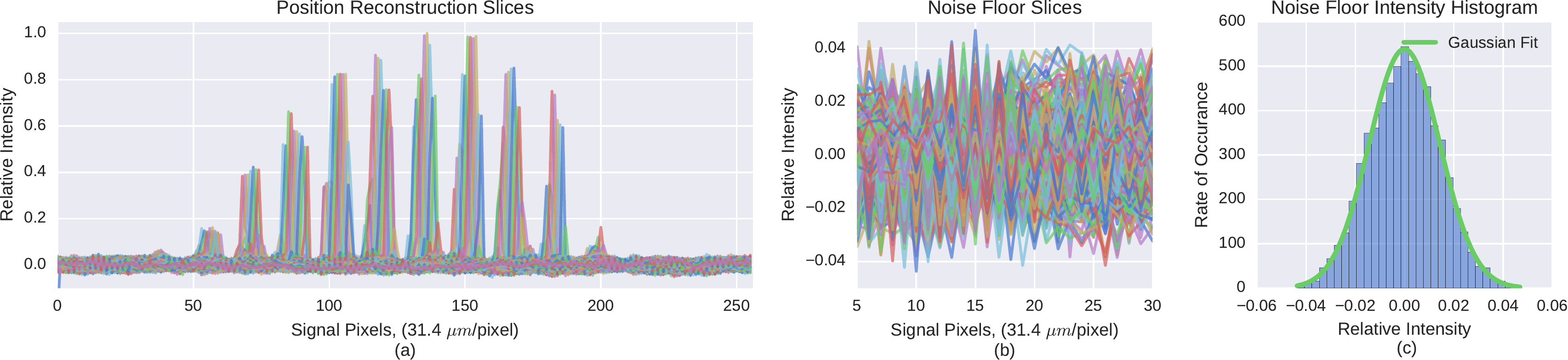}
  \caption{{\color{black} {\bf{Reconstruction noise:}} One-dimensional
      slices along the signal axis of the joint-position
      reconstruction from Fig. \ref{fig:recon} in (a) reveal the
      presence of zero-mean, additive Gaussian noise.  The presence of
      negative values strongly suggests this noise's form is
      non-physical; the reconstruction process maps measurement
      uncertainty into this noise. A close-up of a noise-only region
      (signal pixels 5 to 30, all idler pixel spectra) is shown in
      (b). A histogram of outcomes (c) for the region shown in (b)
      demonstrates that the noise follows Gaussian statistics with
      zero mean and standard deviation .014 . To obtain a valid
      probability distribution, values below a chosen threshold can be
      set to zero and the distribution normalized.}}
  \label{fig:noise}
\end{figure*}

\begin{figure*}
    \includegraphics[scale=0.5]{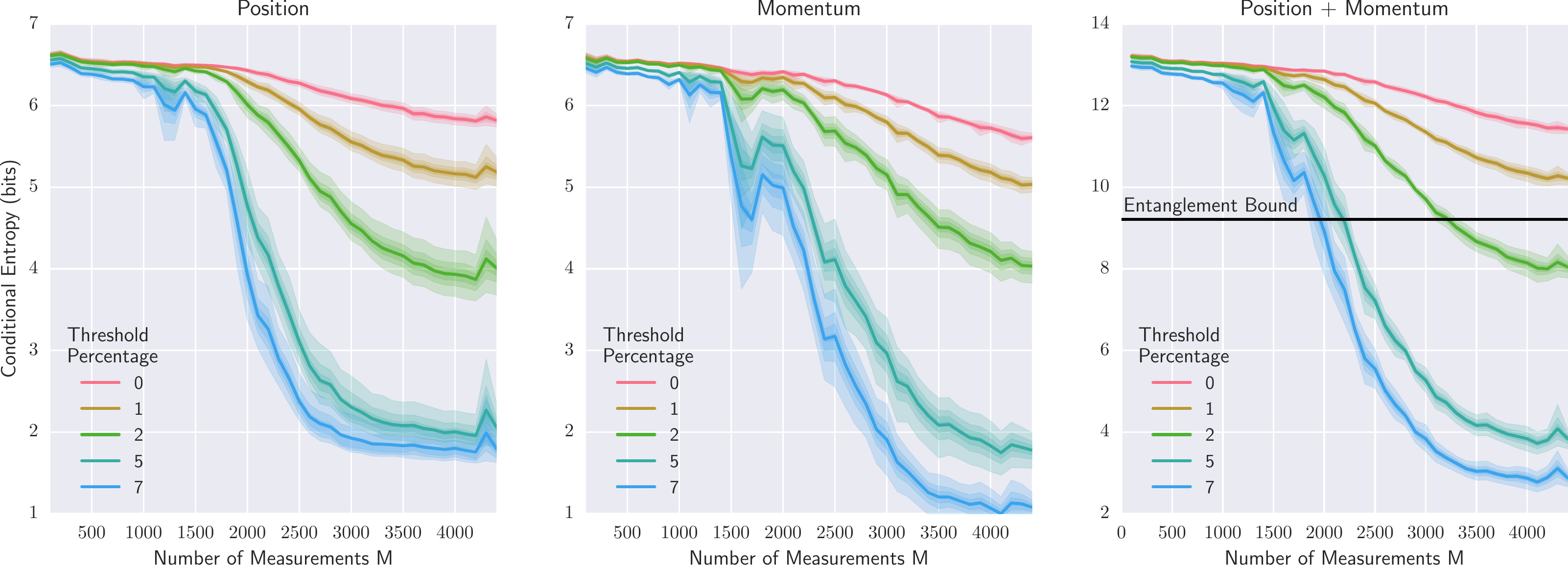}
    \caption{{\bf Conditional entropy versus measurement number:} A
      sharp transition from high to low conditional entropy is seen as
      the number of measurements increases. Note that $N=256^2$, so
      $M=2,000$ is only $.03N$. Different curves correspond to
      different levels of thresholding to remove the noise floor. Bold
      lines indicate an average over for 9 trials.  Faded lines
      enclose up to four standard deviations about the mean. When the
      conditional entropy sum is below the bound, the state is
      entangled.}
    \label{fig:infovsm}
\end{figure*}

{\color{black} \subsection{Signal Recovery}}

Sample recovered joint {\color{black} signals} for position and momentum
are given in Fig. \ref{fig:recon} as returned directly by the
solver. The single-particle resolution was $n=16\times16$ pixels, so
the joint signal has dimensionality $N=n^2=65,536$. For the sample
image, $M=4,439$ random projections were used corresponding to $M$
less than $0.07N$.  Positive correlations in position and negative
correlations in momentum between signal and idler particles are
clearly seen. The gaps visible on the diagonal are an artifact of
row-wise reshaping to one-dimension---these regions are physically
outside the marginal beam width.


{\color{black}
\subsection{Reconstruction Noise}
Unfortunately, the images shown in Fig. \ref{fig:recon} do not
represent valid probability distributions due to the presence of weak,
zero-mean, additive noise shown in Fig. \ref{fig:noise}. Note that the
objective function, Eq. \ref{eq:opt}, does not restrict to valid
probability distributions and allows negative values. We found that
current, established solvers such as TVAL3 performed better without
such additional constraints---improved, quantum-specific solvers are a
topic of future research.

Fig. \ref{fig:noise}(a) shows slices of the joint-position
reconstruction along the signal axis, where each curve corresponds to
a particular idler pixel. Zooming in on a region with no signal in
Fig. \ref{fig:noise}(b), we observe the noise. This noise contains
both measurement uncertainty and solver artifacts.  Potential noise
sources include shot-noise, long term drift in the pump laser, stray
light, and crystal temperature instability. Fig. \ref{fig:noise}(c)
gives a histogram of the noise shown in Fig. \ref{fig:noise}(b) which
follows Gaussian statistics.  An appropriate model for signals
returned by the solver is therefore
\begin{align}
  \bm{X^{(r)}} &= \bm{X} + \bm{G}^{(x)}\\ 
  \bm{K^{(r)}} &= \bm{K} + \bm{G}^{(k)},
\end{align}
where $\bm{X}^{(r)}$ and $\bm{K}^{(r)}$ refer to the signals returned by the solver and $\bm{G}^{(x)}$ and $\bm{G}^{(k)}$ are additive, zero-mean Gaussian noise.

The simplest way to obtain valid probability distributions is to
threshold values below a small percentage of the maximum value to
zero. As seen in Fig. \ref{fig:noise}(b), any threshold below $5\%$
removes the uniform noise floor without removing any signal
peaks. This approach is similar to the common technique of subtracting
dark counts from data in coincidence measurements and other noise
suppression techniques.}

\subsection{Witnessing entanglement}

To witness and quantify entanglement, we violate an entropic steering
inequality \cite{wiseman:2007, walborn:2011, schneeloch:2013:2} (see
supplemental material); all classically correlated states satisfy
\begin{equation}
  H(X_1|X_2)+H(K_1|K_2) \ge 2\log\left(\frac{\pi e}{\Delta_x\Delta_k}\right),
  \label{eq:steer}
\end{equation}
where $H(X_1|X_2)$ and $H(K_1|K_2)$ are the conditional, discrete
Shannon entropies of the respective position and momentum
joint-distributions. {\color{black} $\Delta_k$ ($\Delta_x$) is the
  width in momentum (position) sampled by a single pattern pixel on
  the SLM (DMD) in the transverse plane of the nonlinear crystal. For
  position $\Delta_x$, this is found by dividing the physical width of
  a pattern pixel on the DMD by the magnification of the imaging
  system. For momentum, the physical width of an SLM pattern pixel
  $p_k$ is related to $\Delta_k$ via the Fourier transforming property
  of a lens, so $\Delta_k = p_k2\pi/(\lambda f)$, where $\lambda$ is
  the wavelength of light and $f$ is the lens focal length.}

The entropic steering inequality is powerful because it is computed
directly from measured probability distributions and does not require
a density matrix. Remarkably, despite being a function of discrete
distributions, it witnesses continuous-variable
entanglement. Moreover, the amount the inequality is violated
corresponds to a secret key rate for quantum key distribution
\cite{wiseman:2012,schneeloch:2015:2}.

The conditional entropies in position and momentum for our
experimental results are given in Fig. \ref{fig:infovsm} as a function
of measurement number. Different curves correspond to increased levels
of thresholding, setting values below a percentage of the maximum
value to 0.  A sharp transition from poor reconstruction to good
reconstruction is clearly demonstrated by dramatic drops in the
conditional entropies around $M=2,000$. {\color{black} This transition
  is characteristic of compressed sensing as the number of
  measurements becomes sufficient to accurately reconstruct the signal
  \cite{ganguli:2010}---strongly suggesting we made enough
  measurements. For too small $M$, reconstructions fail spectacularly
  and return unstructured noise. For a $k$-sparse signal ($k$ out of
  $N$ elements have significant intensity), the required number of
  measurements scales as $ck\log(N/k)$ where $c$ is a near-unity
  constant \cite{candes:2006}. For $M$ beyond the transition, one is
  sampling at above the information rate. Traditionally one is
  concerned with sampling at or beyond the Nyquist rate where $M=N$.}

In momentum, the conditional entropy drops to nearly zero; in position
it drops to less than $2$ bits.{\color{black} The position entropy
  likely levels off due to slight pixel-misalignment between the two
  position DMDs. Physically, this indicates a particular signal
  position pixel is correlated to about four idler pixels, whereas a
  particular signal momentum pixel is only correlated to one idler
  pixel.} The steering inequality is violated with as little as $2$
percent thresholding, and by over $6$ bits for thresholding beyond $7$
percent.

\begin{figure}
  \begin{centering}
    \includegraphics[scale=0.5]{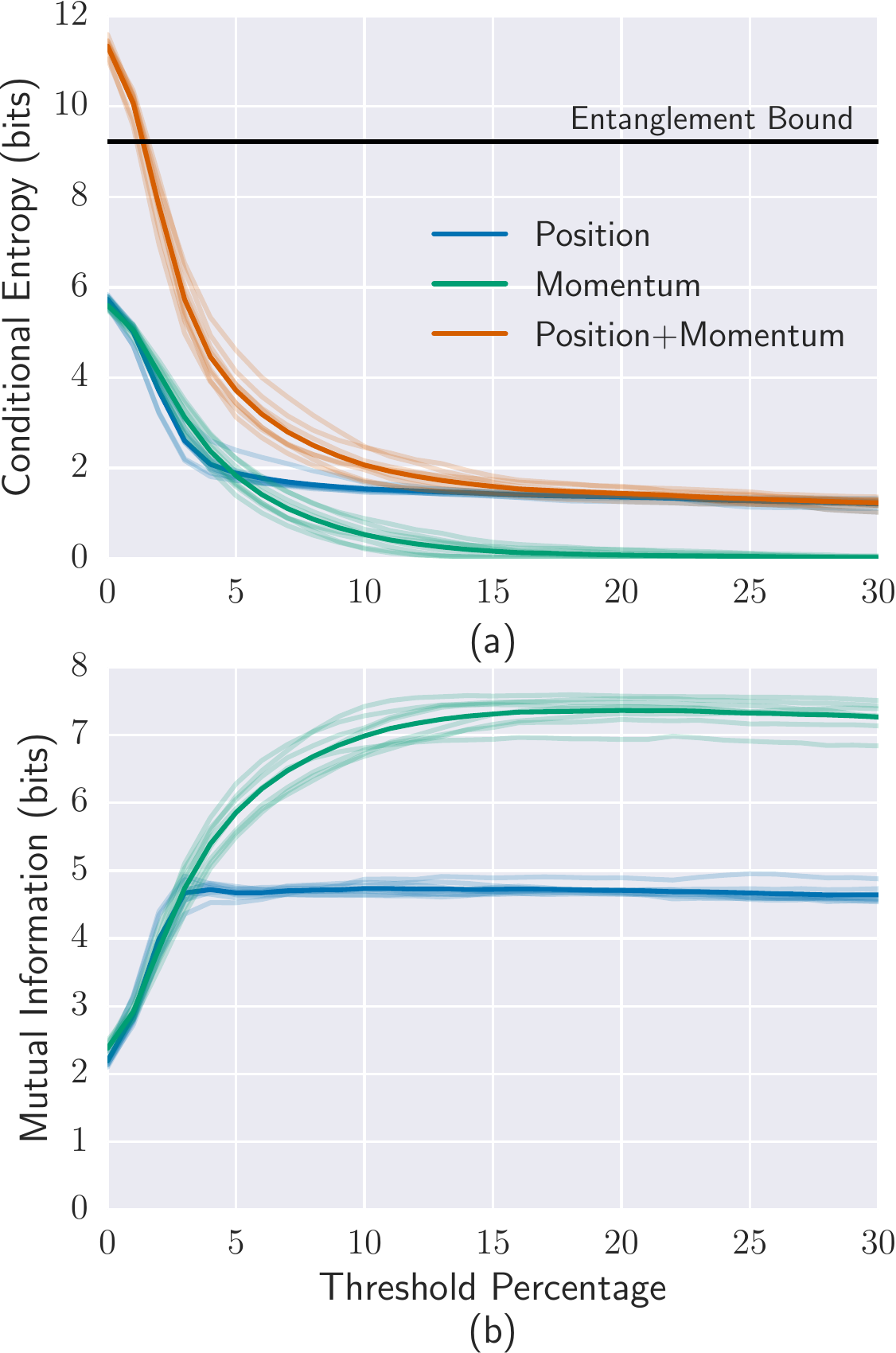}
    \caption{{\bf Effect of threshholding:} The effect of thresholding
      to remove weak background noise on the conditional entropy (a)
      and mutual information (b) is given. The bold line gives the
      average for 9 trials, faded lines give the results from the
      individual trials. $M=4,439$ measurements were used. When the
      conditional entropy sum is below the bound, the state is
      entangled.}
    \label{fig:infovsthresh}
  \end{centering}
\end{figure}

The effect of thresholding for $M=5,000$ is given in
Fig. \ref{fig:infovsthresh}. {\color{black}
  Fig. \ref{fig:infovsthresh}(a) shows the conditional entropies for
  position, momentum, and their sum with the corresponding
  entanglement bound. Fig. \ref{fig:infovsthresh}(b) gives the mutual
  informations $I(X_1:X_2)$ and $I(K_1:K_2)$, where for example
\begin{equation}
  I(X_1:X_2) = H(X_1) + H(X_2) - H(X_1,X_2).
\end{equation}
Here, $H(X_1,X_2)$ is the Shannon entropy of the joint distribution
and $H(X_1)$ and $H(X_2)$ are Shannon entropies of the marginal,
single particle distributions. From information theory, these mutual
informations provide a maximum bit-rate for communication with
joint-position or joint-momentum representations for this system
\cite{cover:2012}. The mutual information rises as a function of
thresholding, indicating that thresholding is not trivially decreasing
the conditional entropies and that the most likely joint-outcomes are
the most highly correlated. Again, the momentum mutual information is
larger due to slight optical misalignments for position DMDs.

An important point is that the thresholded signal peaks still retain
the additive Gaussian noise from the reconstruction process. Due to
the data processing inequality \cite{cover:2012}, this noise cannot
decrease the conditional entropy and cannot increase the mutual
information (this would be like arguing a noisy channel is better for
communication than its noiseless counterpart). Therefore, we
conservatively underestimate our ability to violate the steering
witness (Eq. \ref{eq:steer}).}

\section{Conclusion}
We have demonstrated that local, random filtering in momentum followed
by local, random filtering in position---of the same photons---can
recover sharp, joint distributions for both observables. This is not
possible with standard, projective measurements that localize photons
in either position or momentum. Using the expectation that the signals
will be highly correlated allows us to use many fewer measurements
than dimensions in the system via techniques of compressed sensing. We
strongly emphasize that we have not violated any uncertainty
relations; instead, we have chosen non-projective measurements whose
disturbance can easily be mitigated.

\section{Acknowledgments}

This work was funded by Air Force Office of Scientific Research
(AFOSR) Grant No. FA9550-13-1-0019 and AFOSR LRIR
14RI02COR. G. A. H. and J. S. acknowledge support from National
Research Council Research Associate Programs. J. C. H. acknowledges
support from Northrup Grumman. Any opinions, findings and conclusions
or recommendations expressed in this material are those of the
author(s) and do not necessarily reflect the views of AFRL.

G. A. H. conceived of the experiment and authored the manuscript with
help from S. H. K., D. J. L., and J. S.  S. H. K. and
G. A. H. performed the experiment and analyzed the data.
J. S. provided the theory on entanglement witnesses and Fourier
transforms of random patterns. D. J. L. devised the scheme for using
Hadamard matrices in the measurement and reconstruction process. The
entire project was overseen by J. C. H.  G. A. H. and
S. H. K. contributed equally to the work presented in this manuscript.

\bibliography{refs}

\onecolumngrid
\appendix








\section{Supplemental Material}

\subsection{Fourier transform of  a random, binary filter}

Let $q$ and $p$ be complementary variables where functions in $q$ and $p$ are related by a Fourier transform. $q$ represents the domain being filtered (in our case, momentum). A local, random, $N-$pixel filter in $q$ can be represented as a sum of $n$ top-hat pixel functions arranged on a regular lattice, each multiplied by zero or unity with probability $1/2$;
\begin{equation}
  f_i(q) = \sum_{l}a^{(i)}_l\bigsqcap_{W}(q-W l/2),
\end{equation}
where $W$ is the width of a pixel. Taking the Fourier transform, we obtain
\begin{equation}\label{eq:ft}
  f_i(p) = \frac{W}{\sqrt{2\pi}}\env(p)\times
  \sum_{l}a^{(i)}e^{-iWlp/2},
\end{equation}
where the envelope $\env(p)$ is
\begin{equation}  
  \env(p) = \sinc\left(\frac{W p}{2}\right).
\end{equation}
For small $W$, as with a high resolution pattern in a small window, $\env(p)$ is broad and nearly uniform.

For large $N$, the summation term in Eq. (\ref{eq:ft}) can be
modeled as a sum of $N/2$ phasors of unit length. At $p=0$, the
phasors add in phase so the $p=0$ component is $N/2$. Because the
nonzero elements of $a^{(i)}$ are randomly distributed, values of
$f(p)$ for $p\ne 0$ can be modeled as a sum of $N/2$ randomly oriented
unit-length, phasors modulated by $\env(p)$. This sum is effectively a
two-dimensional random walk in the complex plane. For large values of
$N$ typical in imaging, the resulting sampling distribution is a
circularly-symmetric Gaussian distribution in the complex plane.

A model for $f(p)$ is therefore,
\begin{equation}
  f(p) = \frac{N}{2}\env(p)\left(\delta(p) + \sqrt{\frac{2}{N}}\phi_i(p)\right),
\end{equation} 
where values for $\phi_i(p)$ are complex numbers with uniformly varying phase and square magnitude distributed according to a $\chi^2$ distribution with standard deviation $\sigma=\sqrt{1/2}$ so $\mean{|\phi_i(p)|^2}=1$ and $\mean{\cdot}$ is an average over many filter functions.

The end result is effectively a large delta function at $p=0$ riding a small noise floor a factor $\sqrt{N/2}$ weaker, all modulated by a envelope function given by the width of a pixel in $q$. 

\subsection{Effect of random, partial projections in the complementary domain}

Consider filtering a bipartite wavefunction $\psi(q_1,q_2)$ with random local filters $f_i(q_1)$ and $g_i(q_2)$, which is the product
\begin{equation}
  \tilde{\psi}(q_1,q_2)=\psi(q_1,q_2)f_i(q_1)g_i(q_2).
  \label{eq:filt}
\end{equation}
The effect of filtering in the complementary domain is found by taking the Fourier transform of Eq. \ref{eq:filt}. Because the product of two functions in one domain is a convolution in the complementary domain, we find
\begin{align}
  \tilde{\psi}(p_1,p_2) &= \mathcal{F}\left\{\psi(q_1,q_2)f_i(q_1)g_i(q_2) \right\}\\
& = \mathcal{N}\psi(p_1,p_2)\star\left[\left(\delta(p_1) + \sqrt{2/N}\phi_i(p_1)\right)\left(\delta(p_2)+\sqrt{2/N}\phi_i(p_2)
\right)\right],
\end{align} 
where $\mathcal{N}$ is a normalization constant, envelope functions $\env(p)$ are assumed to be uniform, and $\star$ denotes convolution. Because convolution with a $\delta$-function returns the original function, this result will give the true momentum distribution plus a series of weak, additive noise terms. It is convenient to expand in powers of $1/\sqrt{N}$ to give
\begin{align}
|\tilde{\psi}(p_1,p_2)|^{2}&=\mathcal{N}\bigg[|\psi(p_1,p_2)|^{2}\nn\\
&+ \frac{2\sqrt{2}}{\sqrt{N}}\mathbf{Re}[\psi^{*}(p_1,p_2)(\psi(p_1,p_2)\star(\delta(p_1)\phi_{j}(p_2))+\delta(p_2)\phi_{i}(p_1)))]\nn\\
&+ \frac{4}{N}\mathbf{Re}[\psi^{*}(p_1,p_2)(\psi(p_1,p_2)\star(\phi_{i}(p_1)\phi_{j}(p_2))]\nn\\
&+\frac{2}{N}|\psi(p_1,p_2)\star(\delta(p_1)\phi_{j}(p_2))+\delta(p_2)\phi_{i}(p_1))|^{2}\nn\\
&+ \frac{2^{\frac{5}{2}}}{N^{\frac{3}{2}}}\mathbf{Re}\big[(\psi^{*}(p_1,p_2)\star(\delta(p_1)\phi^{*}_{j}(p_2))+\delta(p_2)\phi^{*}_{i}(p_1)))*\nn\\
&*(\psi(p_1,p_2)\star\phi_{i}(p_1)\phi_{j}(p_2))\big]\nn\\
&+\frac{4}{N^{2}}|\psi(p_1,p_2)\star\phi_{i}(p_1)\phi_{j}(p_2)|^{2}
\bigg],
\end{align}

When averaging over many patterns, coherent interference terms average to zero, yielding the simpler expression
\begin{align}
\langle|\tilde{\psi}(p_1,p_2)|^{2}\rangle & \approx\mathcal{N'}\bigg[|\psi(p_1,p_2)|^{2}\nn\\
&+\frac{2}{N}|\psi(p_1,p_2)\star(\delta(p_1)\phi_{j}(p_2))+\delta(p_2)\phi_{i}(p_1))|^{2}\nn\\
&+\frac{4}{N^{2}}|\psi(p_1,p_2)\star\phi_{i}(p_1)\phi_{j}(p_2)|^{2}
\bigg],
\end{align}
where $\langle\cdot\rangle$ is an average over many filter functions.

\subsection{Full measurement process}

In order to measure the interaction of the light with a single set of
random patterns, one needs $16$ different coincidence measurements
corresponding to all combinations for the transmitting and rejection
ports for each filter. These can be performed simultaneously with $8$
detectors. However, because we had only two detectors, we performed
them in sequence. These measurements must be combined to decouple
position from momentum and fit the linear measurement model.

\begin{align}
  \bm{Y}^{(k)} &= \bm{AK} + \bm{\Phi}^{(k)}\\
  \bm{Y}^{(x)} &= \bm{BX} + \bm{\Phi}^{(x)} + \bm{\Gamma}^{(x)}.
  \label{eq:meas}
\end{align}
Here, $\bm{X}$ and $\bm{K}$ are $N$-dimensional signal vectors
representing $|\psi(x_1,x_2)|^2$ and $|\psi(k_1,k_2)|^2$, $\bm{A}$ and
$\bm{B}$ are $M \times N$ sensing matrices, $\bm{\Phi}$ are
$M$-dimensional noise vectors. $\bm{\Gamma}^{(x)}$ is the extra noise
injected into the $\bm{X}$ signal by first filtering in
$\bm{K}$. $\bm{Y}^{(k)}$ and $\bm{Y}^{(x)}$ are measurement vectors
whose elements are inner-products of $\bm{X}$ or $\bm{K}$ with the
$i^{\text{th}}$ row (or sensing vector) of $\bm{A}$ or
$\bm{B}$. $\mathbf{\Gamma}^{(x)}$ is a noise vector representing the
noise introduced as a consequence of filtering in momentum.

An $N$-dimensional sensing vector, e.g. $\bm{A}_i$, consists of the
Kronecker product of two $n$-dimensional, single-particle sensing
vectors such that
$\bm{A}_i = \bm{a}_i^{(k_1)}\otimes \bm{a}_i^{(k_2)}$. The elements of
$\bm{a}_i^{(k_1)}$ and $\bm{a}_i^{(k_2)}$ randomly take values ``1''
(transmit) and ``-1'' (reject) with equal probability. A similar
process defines single-particle sensing vectors $\bm{b}$ in position.

There are therefore $16$ possible filter combinations for a given set
of $\bm{a}_i$ and $\bm{b}_i$ yielding correlation measurements $y_1$
through $y_{16}$;
\begin{equation}
  \begin{blockarray}{cccccccc}
    & & \BAmulticolumn{4}{c}{\text{Momentum Filters}} & &\\
    & & \BAmulticolumn{4}{c}{\overbrace{\phantom{qqqqqqqqqqqqqqqqq}}} & &\\
    & & \mathcal{TT} & \mathcal{TR} & \mathcal{RT} & \mathcal{RR} & &\\ 
 \begin{block}{\BAmultirow{0.5in}\{l(cccc)@{\hspace{10pt}}\BAmultirow{0.4in}(c)}
      \text{Position}& \phantom{y}\mathcal{TT}\phantom{yy} & y_1 & y_2 & y_3 & y_4 & & B^{\mathcal{TT}} \\
      \text{Filters}&\phantom{y}\mathcal{TR}&y_5&y_6&y_7&y_8& $\sum\limits_{\text{cols}}=$ &B^{\mathcal{TR}}\\
      &\phantom{y}\mathcal{RT}&y_9&y_{10}&y_{11}&y_{12}& &B^{\mathcal{RT}}\\
      &\phantom{y}\mathcal{RR}&y_{13}&y_{14}&y_{15}&y_{16}& &B^{\mathcal{RR}}\\
    \end{block}
    & & \BAmulticolumn{4}{c}{\sum\limits_{\text{rows}}=} & & \\
    \begin{block}{cc(cccc)cc}
      & & A^{\mathcal{TT}} & A^{\mathcal{TR}} & A^{\mathcal{RT}} & A^{\mathcal{RR}} &. &\\
    \end{block}
  \end{blockarray}
\end{equation}

By summing over rows (position) or columns (momentum), we can separate
the position and momentum measurements up to the effect of measurement
disturbance, which we have previously established is just a small
amount of Gaussian noise---this noise is included in the model via
$\bm{\Phi}^{(k)}$ and $\bm{\Phi}^{(x)}$. By summing outcomes of
like-acting filters ($\mathcal{TT}$ and $\mathcal{RR}$) and taking the
difference of opposing filters ($\mathcal{TR}$ and $\mathcal{RT}$),
measurement values $Y_i^{(k)}$ and $Y_i^{(x)}$ corresponding to $A_i$
and $B_i$ are generated.
\begin{align}
  \bm{Y}_i^{(k)} &= \frac{\bm{A}_i^{\mathcal{TT}} -\bm{A}_i^{\mathcal{RT}} - \bm{A}_i^{\mathcal{TR}} + \bm{A}_i^{\mathcal{RR}}}{\bm{A}_i^{\mathcal{TT}} +\bm{A}_i^{\mathcal{RT}} + \bm{A}_i^{\mathcal{TR}} + \bm{A}_i^{\mathcal{RR}}} = \bm{A}_i\bm{K} + \bm{\Phi}_i^{(k)}\\
  \bm{Y}_i^{(x)} &= \frac{\bm{B}_i^{\mathcal{TT}} - \bm{B}_i^{\mathcal{RT}} - \bm{B}_i^{\mathcal{TR}} + \bm {B}_i^{\mathcal{RR}}}{\bm{B}_i^{\mathcal{TT}} + \bm{B}_i^{\mathcal{RT}} + \bm{B}_i^{\mathcal{TR}} + \bm {B}_i^{\mathcal{RR}}} = \bm{B}_i\bm{X} +\bm{\Phi}_i^{(x)}\nonumber
\end{align}
The values in the denominator normalize each measurement to the total
coincidences, such that each term represent the probability a
particular detection event occurs in each of the four
possibilities. It also helps compensate for any drift in the the
total, joint intensity over the course of the experiment.

This process is repeated $M$ times to build the full measurement
vectors $\bm{Y}^{(k)}$ and $\bm{Y}^{(x)}$.

\subsection{Simulations with Noise}

\begin{figure}[h]
  \includegraphics[scale=0.5]{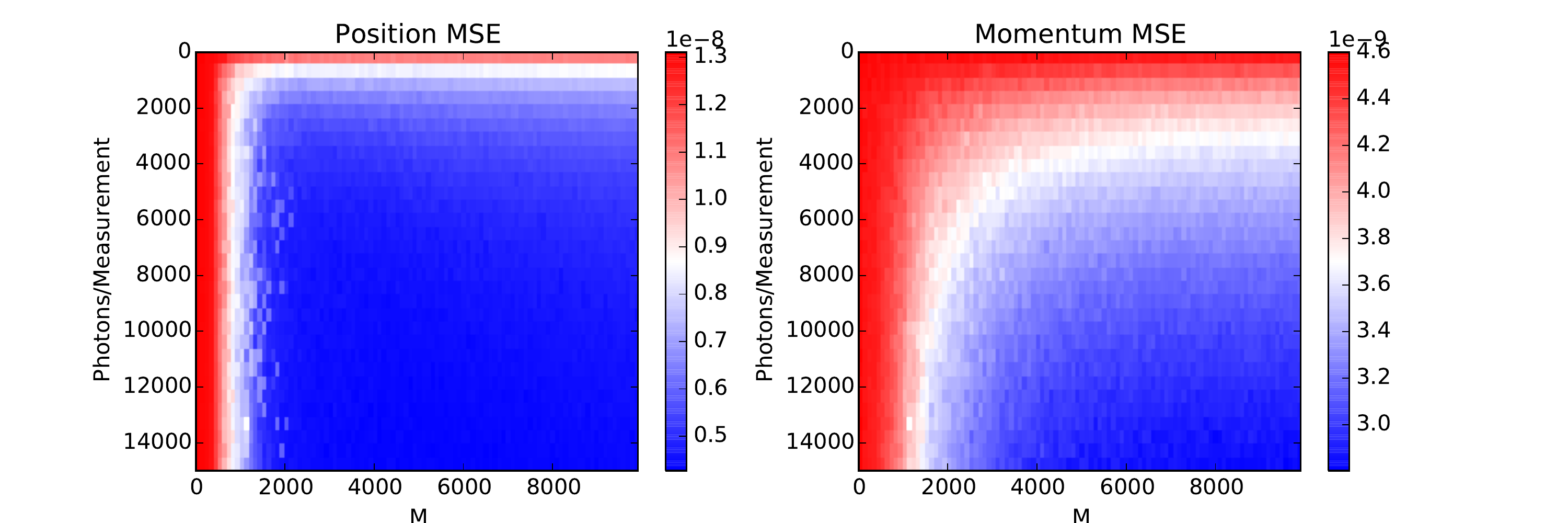}
  \caption{Simulated MSE in momentum and position as a function of $M$
    and average detection events per pattern.}
  \label{supp:fig:noise}
\end{figure}

To investigate the accuracy of our technique, we simulated our
technique as a function of measurement number $M$ and average detected
flux. For the SPDC state, we use the standard double-Gaussian model
\cite{schneeloch:2015}
\begin{equation}
  \ket{\psi} \propto \int dx_1dx_2 e^{-\frac{(x_1-x_2)^2}{8\sigma_-^2}}e^{\frac{(x_1+x_2)^2}{16\sigma_p^2}}\ket{x_1}\ket{x_2}, 
\end{equation} 
where $\sigma_p$ is the transverse standard deviation of the pump laser width and
\begin{align}
  \sigma_- = \sqrt{\frac{9L_z\lambda_p}{20\pi}}.
\end{align}
Here, $L_z$ is the length of the nonlinear crystal and $\lambda_p$ is
the wavelength of the pump laser. For the simulation, we chose
$L_z = 1$ mm, $\lambda_p = 400$ nm, and $\sigma_p = .85$ mm. The state
was sampled above the Nyquist limit with a window including $3$
standard deviations of the double Gaussian in both position and
momentum. For computational simplicity, we simulated in one dimension
per particle. The state was first filtered in position and then
momentum (opposite of the experiment but conceptually identical) by
$M$, $16$ pixel local binary filters for a $16\times 16$ dimensional
joint state.

The mean-square-error (MSE) of the simulation results with respect to
the true distributions in shown in
Fig. \ref{supp:fig:noise}. Compressive sensing reconstruction
techniques typically display a sharp phase-shift from poor quality to
good quality reconstruction as a function of $M$ and photon flux
\cite{donoho:2011}. This phase change can be clearly seen in the
results as the white transition region.

The effect of filtering position introduces extra Gaussian noise into
the the momentum distribution. This has the effect of shifting the
phase transition curve to require slightly higher flux than position
$~4000$ photons/measurement instead of $1000$ and to require slightly
more measurements, which varies depending on flux. Because total
variation minimization (our solver) is effectively a Gaussian
denoiser, it removes the noise injected by our filtering so long as we
modestly increase flux and measurement number.

\subsection{EPR Steering}

EPR Steering is a generalization of the EPR paradox describing
non-local correlation stronger than entanglement, but weaker than
Bell-nonlocality \cite{wiseman:2007}. Practically, the extent to which
a quantum state is steerable is relevant for security in quantum key
distribution \cite{wiseman:2012}. The term steering, coined by
Schr\"{o}dinger, refers to the idea that by choosing a measurement
basis, Alice can ``steer'' Bob's state into an eigenstate of the basis
chosen by Alice.

Steering inequalities follow from uncertainty relations---a
two-particle quantum state is steerable if its conditional variances
in complementary observables are smaller than what the Heisenberg
uncertainty relation allows for unconditioned variances. For example,
given position and momentum, if
$\sigma_{(x_1|x_2)}\sigma_{(k_1|k_2)}<1/2$, the state is steerable.

Schneeloch \emph{et~al.}\cite{schneeloch:2013} showed that all
classically correlated states satisfy an entropic steering inequality;
\begin{equation}
  H(X_1|X_2)+H(K_1|K_2) \ge 2\log\left(\frac{\pi e}{\Delta_x\Delta_k}\right).
  \label{eq:steer}
\end{equation}
$H(X_1|X_2)$ and $H(K_1|K_2)$ are the discrete, conditional entropies
for the binned probability distributions in the position and momentum
of particle $1$, conditioned on measurements of particle
$2$. $\Delta_x$ and $\Delta_k$ refer to the respective position or
momentum discretization widths, and the factor of $2$ in front of the
logarithm on the right-hand side accounts for the transverse position
and momentum each having two dimensions.  Conceptually,
Eq. (\ref{eq:steer}) states that \emph{strong correlations in
  measurements of complementary observables is a signature of
  entanglement}---essentially a restatement of the EPR paradox. Note
that the conditional entropies are calculated directly from the
measured probability distributions, and not from an inferred quantum
state. Despite the fact that the measured distributions are discrete,
Eq. (\ref{eq:steer}) witnesses \emph{continuous-variable} steering.









\end{document}